\documentclass[aps,
twocolumn,
superscriptaddress,
amssymb,
amsmath,
prb,
floatfix,
longbibliography]{revtex4-2}
\usepackage{graphicx }
\usepackage[]{hyperref}
\usepackage{dcolumn}
\usepackage{amsmath}
\usepackage{amssymb}
 \usepackage{amsfonts}
 \usepackage{bbm}

\usepackage{physics}
\usepackage[normalem]{ulem}

\hypersetup{colorlinks=true,linkcolor=blue,citecolor=blue,urlcolor=blue,pdfpagemode=UseNone}

\newcommand{\slrrtext}  {spin-lattice-relaxation rate}
\newcommand{\slrr}      {$T_1^{-1}$}
\newcommand{\VT}[1]{{#1}}

\begin{document}
\title{Observation of a pronounced Hebel-Slichter peak in the spin-lattice relaxation rate and implications for gap and pairing symmetry in LaNiGa$_2$}

\author{P. Sherpa}
    \affiliation{Department of Physics and Astronomy, University of California, Davis, California
95616, USA}

\author{R. Hingorani}
    \affiliation{Department of Physics and Astronomy, University of California, Davis, California
95616, USA}

\author{A. Menon}
    \affiliation{Department of Physics and Astronomy, University of California, Davis, California
95616, USA}

\author{I. Vinograd}
    \affiliation{Department of Physics and Astronomy, University of California, Davis, California
95616, USA}

\author{C. Chaffey}
    \affiliation{Department of Physics and Astronomy, University of California, Davis, California
95616, USA}

\author{A. P. Dioguardi}
    \affiliation{Materials Physics and Applications - Quantum, Los Alamos National Laboratory, Los Alamos, New Mexico 87545, USA}

\author{R. Yamamoto}
    \altaffiliation{Current address: Department of Physics and Astronomy, University of California Los Angeles, Los Angeles, California 90095, USA}

    \affiliation{Materials Physics and Applications - Quantum, Los Alamos National Laboratory, Los Alamos, New Mexico 87545, USA}

\author{M. Hirata}
\affiliation{Materials Physics and Applications - Quantum, Los Alamos National Laboratory, Los Alamos, New Mexico 87545, USA}

\author{F. Ronning}
\affiliation{Materials Physics and Applications - Quantum, Los Alamos National Laboratory, Los Alamos, New Mexico 87545, USA}

\author{J. R. Badger}
    \affiliation{Department of Chemistry, University of California, Davis, California
95616, USA}

\author{P. Klavins}
    \affiliation{Department of Physics and Astronomy, University of California, Davis, California
95616, USA}

\author{R. R. P. Singh}
    \affiliation{Department of Physics and Astronomy, University of California, Davis, California
95616, USA}

\author{V. Taufour}
    \affiliation{Department of Physics and Astronomy, University of California, Davis, California
95616, USA}

\author{N. J. Curro}
    \affiliation{Department of Physics and Astronomy, University of California, Davis, California
95616, USA}
\date{\today}
\begin{abstract}

We report a pronounced Hebel-Slichter coherence peak in the zero field nuclear quadrupolar resonance (NQR) spin-lattice relaxation rate of the topological crystalline superconductor LaNiGa$_2$ in the superconducting state. Previously, a two-band internally antisymmetric non-unitary triplet pairing (INT) state was proposed for this system, with equal spin-pairing and two distinct gaps associated with different spins. A detailed examination of the temperature dependence of the NQR data shows that the data best fit an INT model if the two gaps are equal and the model is unitary. Even a tiny non-unitarity with two unequal gaps causes the coherence peak to diminish rapidly and deviate from the data.
On the other hand, the data are well-fit by a two-band singlet BCS-like pairing with two distinct gaps consistent with previous measurements.  This raises doubts on the identification of {\it non-unitary} triplet-pairing with time-reversal symmetry breaking in this material.

\end{abstract}

\maketitle

Superconductors that exhibit spontaneous magnetic fields are unusual because the most energetically favorable state generally involves spin-singlet pairing between Cooper pairs of opposite momentum electrons.  Spin triplet pairing can emerge in rare cases, such as in the well-known candidates UPt$_3$ and UTe$_2$, where triplet pairing has been observed, and, in case of UPt$_3$  there is further evidence of  a finite magnetization and time-reversal symmetry breaking (TRSB) \cite{Ran2019,Schemm2014}. 
TRSB has been recently observed via muon spin resonance ($\mu$SR) experiments in the superconducting state of the topological crystalline superconductor LaNiGa$_2$ with $T_c = 2.1$ K \cite{Hillier2012,Ghosh2020}. This material has a non-symmorphic orthorhombic crystal structure ($C_{mcm}$) that gives rise to a non-trivial band topology  with band-degeneracies at the Fermi level \cite{Badger2022, Quan2022}.  This degeneracy can give rise to multi-band superconductivity involving both intra- and inter-band pairing. It has been proposed that the condensate wavefunction may be described as an internally antisymmetric, nonunitary, triplet (INT) state that is 
fully gapped with s-wave symmetry in momentum space,  antisymmetric in the band channel, and symmetric in spin space with equal-spin triplet ($S=1$) pairing \cite{Quintanilla2020}.  In this case, there are two components to a superconducting pair wavefunction corresponding to either both spins up or both spins down \cite{Weng2016,Quintanilla2020}. If the order parameter is nonunitary, there will be  two distinct superconducting gaps with different magnitudes, and the corresponding superfluid densities will also differ leading to a finite internal magnetization. 
Support for this interpretation has emerged from specific heat \cite{Weng2016}, penetration depth  \cite{Ghimire2024}, and $\mu$SR \cite{Sundar2024} measurements in LaNiGa$_2$ that revealed the presence of two superconducting gaps. However, such  measurements are mostly sensitive to the density of states and cannot probe the spin structure of the superconducting condensate.  

\begin{figure}
\centering
\includegraphics[width=\linewidth]{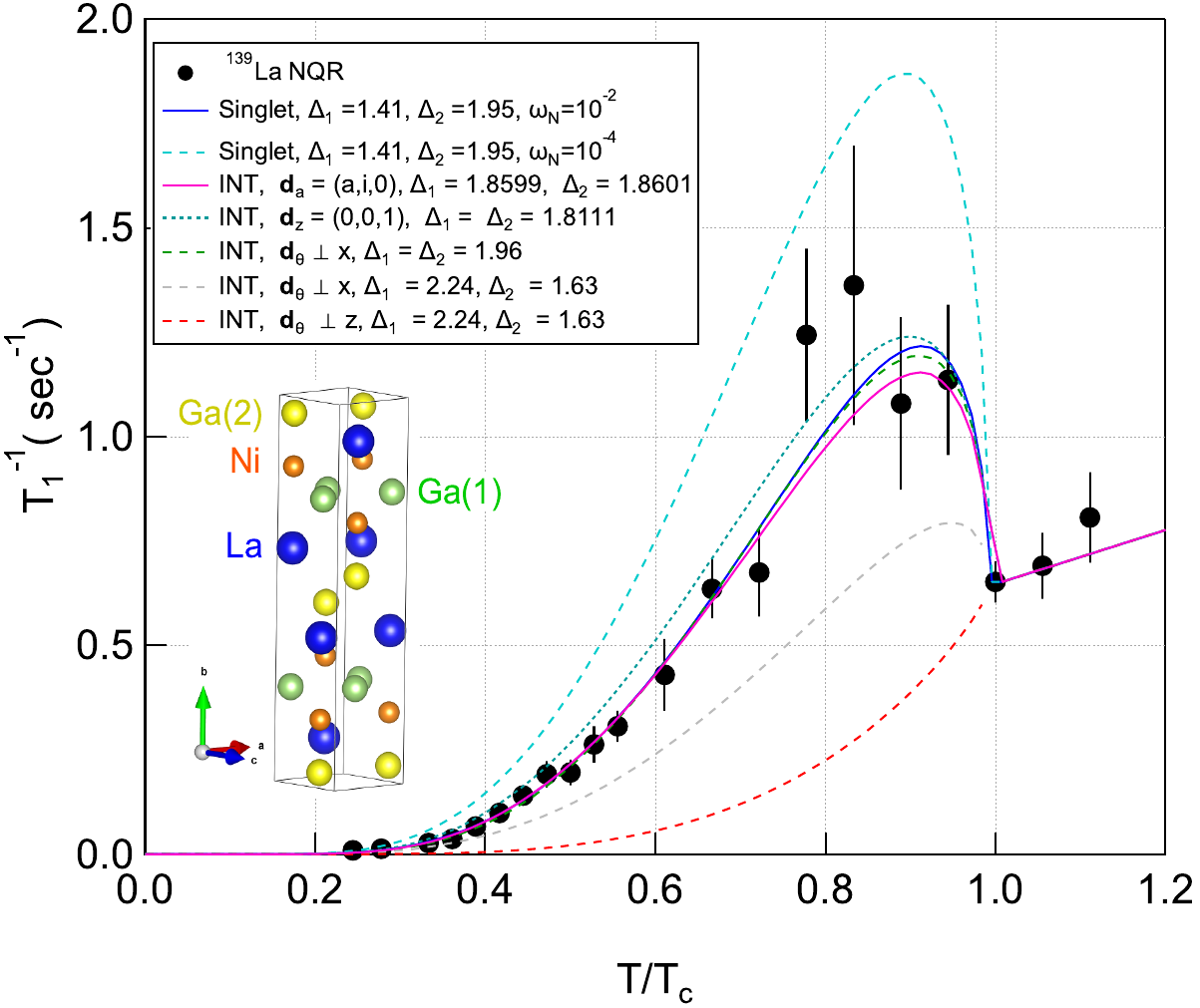}
\caption{\label{Fig:T1inv} $^{139}$La \slrr\ versus inverse temperature in the superconducting state of LaNiGa$_2$ (with $T_c = 1.83$ K).  The solid and dashed lines are calculations as discussed in the text, {and the gap values, ($\Delta_1$, $\Delta_2$) and nuclear resonance frequency, $\omega_N$, are given in units of $k_BT_c$}. 
The data are well fit by a two-gap singlet-pairing model as well as by an INT triplet-pairing model with two equal gaps, but not with the INT model with two different gaps.  The inset shows the unit cell of LaNiGa$_2$. }
\end{figure}

Zero field nuclear quadrupolar resonance (NQR) can shed light on the spins of Cooper pairs because the spin lattice relaxation rate, \slrr,  of nuclear spins is driven by spin-flip scattering with the Bogoliubov excitations.
The Hebel-Slichter coherence peak is an increase in \slrr\ below $T_c$ that arises due to a temperature-dependent build up of a sharp peak in the density of states and the particular quantum mechanical superposition of holes and electrons describing the excitations  \cite{hebelslichter1,hebelslichter2,ArovasSuperconductivity}. The coherence peak generally emerges in conventional superconductors \cite{maclaughlinSCreviewbook,Kiefl1993,MgB2NMR}, but is suppressed, apart from extrinsic impurity scattering, for several intrinsic reasons in unconventional superconductors. These include an anisotropic gap that can suppress the peak in the density of states or a change in the symmetry of the pair-wave functions and excitations that can alter coherence factors \cite{Statt1990,Brandow1998}. Here we find that in an equal-spin pairing triplet state the peak can also be suppressed by having two different gaps for different spin components. At lower temperatures, the temperature dependence of \slrr\ reflects the superconducting gap, exhibiting either activated behavior in superconductors with a full gap \cite{IshidaLaFe4P12NMR,Ding2016}, or power law behavior in unconventional superconductors with nodes in the gap \cite{imaiYBCOT1,Curro2005}.  Less is known about the behavior of \slrr\ for the INT state. In this case, there are four types of Bogoliubov excitations involving superpositions of electrons and holes with parallel spins in different {bands}. Because they involve parallel spins, one might expect spin-flip scattering by nuclear spins would be suppressed, and that \slrr\ may depend on the orientation of the $\mathbf{d}$-vector that describes the triplet order parameter \cite{NMRchiral2016,T1andT2chiralSC}.  

Here we present $^{139}$La ($I=7/2$) NQR \slrr\ data in the superconducting state of LaNiGa$_2$ that exhibits a Hebel-Slichter coherence peak as well as activated behavior at low temperatures.  We show that a robust coherence peak can only emerge for the INT state if the two gaps are equal in magnitude, with no TRSB. 
When the equal spin pairing axis is different from the nuclear quantization axis, a weak HS peak arises with TRSB with two distinct gaps,
but it is too weak to fit the data.
Comparison of the observed HS peak in LaNiGa$_2$ with other materials shows that the peak in LaNiGa$_2$ is quite robust.
Our experimental results are well described by a two-band singlet model with different gap magnitudes consistent with the previously reported heat capacity and penetration depth experiments.

\begin{figure}
\centering
\includegraphics[width=\linewidth]{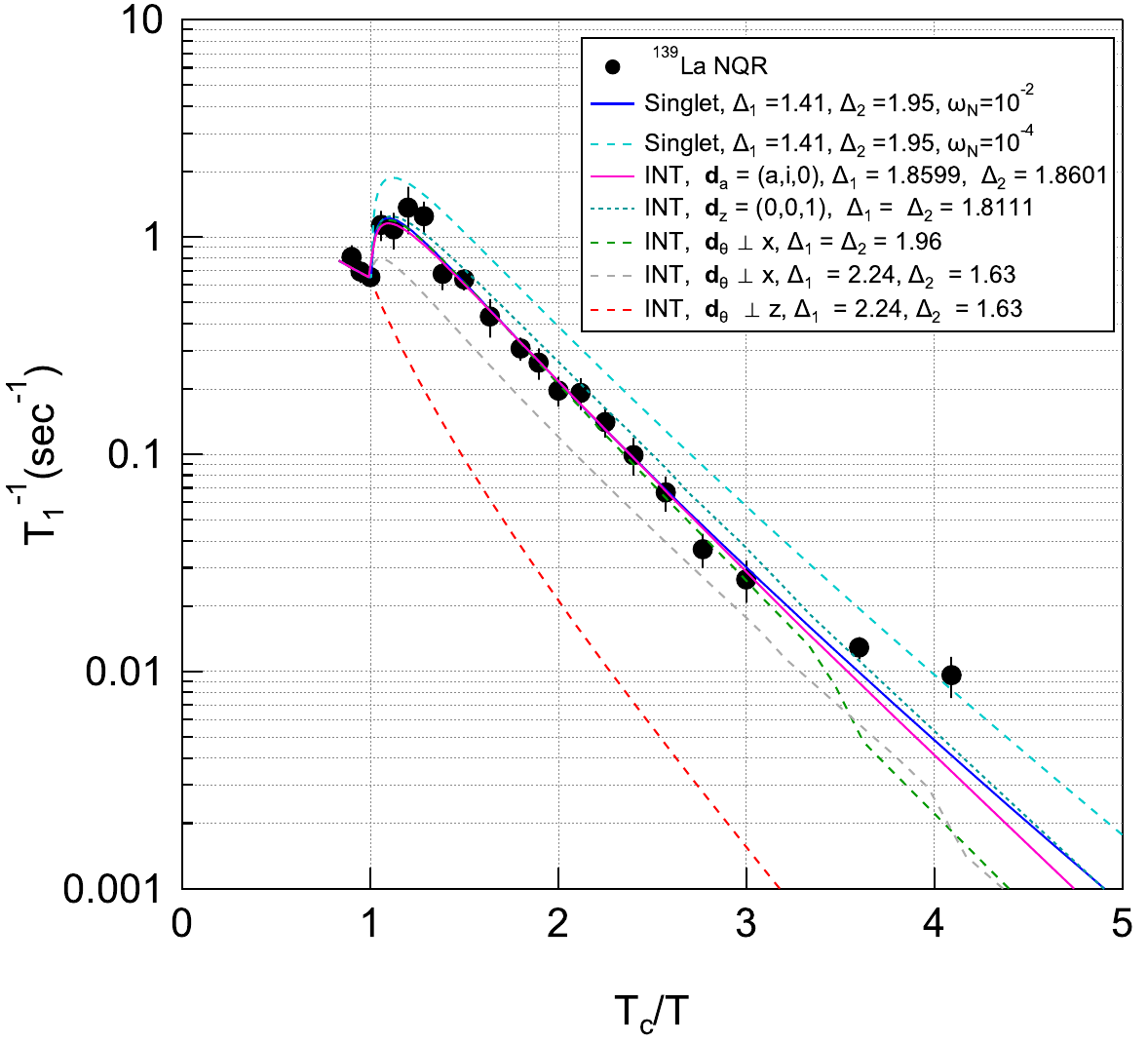}
\caption{\label{Fig:T1invlog} $^{139}$La \slrr\ versus inverse temperature in the superconducting state.  The solid and dashed curves are calculated values as discussed in the text, {and the gap values are given in units of $k_BT_c$}.  }
\end{figure}

$^{139}$La NQR measurements of the spin lattice relaxation rate were carried out on a finely powdered sample of LaNiGa$_2$ ($T_c = 1.83$ K) down to 240 mK, as discussed in the supplemental materials \cite{supp}.  This material has a single La site (see inset of Fig. \ref{Fig:T1inv}), and the quantization axis of the $I=7/2$ nuclear spins lies along the principal axis of the electric field gradient (EFG) tensor, $\nu_{\alpha\beta}$. This axis corresponds to the crystalline $b$ axis with $\nu_{bb} = 1.65$ MHz, and asymmetry parameter $\eta=(\nu_{aa} - \nu_{cc})/\nu_{bb} = 0.04$ \cite{Sherpa2024}.  \slrr\  measurements were carried out at the $3\nu_Q \approx 5.060$ MHz transition by inversion recovery, and the magnetization recovery was fit to the standard expression for magnetic fluctuations.  The measurements were performed in a top-loading dilution refrigerator, with one fixed and one variable capacitor that can be adjusted at room temperature.  The sample was immersed in the liquid helium in the mixing chamber, and the radiofrequency power was reduced to suppress spurious heating of the sample.  Tests were conducted in which \slrr\ was measured for several different power levels, and we observed little to no change in the value of \slrr\ 
with power levels of 30 W and pulse widths $\sim 30$ $\mu$s. Figure \ref{Fig:T1inv} shows the temperature dependence of \slrr\ in the superconducting state.  There is a clear enhancement of the relaxation rate below $T_c$ by about a factor of 2.5, and at lower temperatures \slrr\ is suppressed exponentially, as shown in Fig. \ref{Fig:T1invlog}.  There is a change in the slope around $T_c/T \approx 3.5$, suggesting the presence of multiple gaps, with values of approximately $2 k_BT_c$ and $< 1.0 k_B T_c$ estimated from the slopes.
It is possible that an impurity band may be giving rise to residual scattering at lowest temperatures \cite{Curro2005}.

\begin{table}[b]
\caption{\label{tab:gaps}Superconducting gap values measured in LaNiGa$_2$.}
\begin{ruledtabular}
    \begin{tabular}{lll}
        source & $\Delta_1/k_BT_c$ & $\Delta_2/k_BT_c$ \\ \hline
        NQR singlet fit & 1.41 & 1.95 \\
        NQR triplet fit ($\textbf{d}_{\theta}$) & 1.96 & 1.96 \\ 
        NQR triplet fit ($\textbf{d}_a$) & 1.8599 & 1.8601 \\
        NQR triplet fit ($\textbf{d}_z$) & 1.81 & 1.81 \\
        heat capacity \cite{Weng2016}  & 1.29 & 2.04 \\
        susceptibility \cite{Ghimire2024} & 1.57 & 1.89 \\  
        $\mu$SR \cite{Sundar2024}  & 1.3 & 2.7 \\
\end{tabular}
\end{ruledtabular}
\end{table}

To compute the superconducting gap and excitations, we ignore details of the band structure and consider two different Bogoliubov-de-Gennes (BdG) Hamiltonians of similar essence with isotropic gap, inspired by the INT model introduced in \cite{Ghosh2020,Quan2022}. These two models differ in the pairing mechanism: the INT state has equal-spin triplet pairing, whereas the second has singlet pairing and is a two-gap BCS model. Both have two bands  with inter-band pairing. The INT case is antisymmetric in the  band index with parallel spin, but with an admixture of up-up and down-down which allows TRSB and two gaps. The BCS case is symmetric in the band index but has spin singlet. Both models still allow two gaps, and in either case, the gap is assumed to be isotropic (the spatial
wave-function is s-wave). We are, thus, primarily exploring two gaps with the associated build-up of  density of states and the nature of coherence factors that come from different pairing symmetries.

We note that the triplet states of the INT model are very different from the BW (Balian-Werthamer) \cite{BW} and ABM (Anderson-Brinkman-Morel) \cite{AM} states
proposed for triplet superfluids and realized in the B and A phases of $^3$He. In that case there is no band structure, and the anti-symmetry comes from real space wave-functions, which are \textbf{k}-dependent and odd in \textbf{k}. 
{In LaNiGa$_2$ with $D_{2h}$ point group symmetry, the allowed irreducible representations for a single-band spin triplet model with weak spin orbit coupling all show line or surface nodes in the gap for spherical or cylindrical Fermi surfaces~\cite{Annett1990AP}.}

In order to model the spin lattice relaxation in LaNiGa$_2$, we assume that the La nuclear spins interact with the electron spins in two different bands via a contact hyperfine interaction:  $\mathcal{H}_{hyp} = A \mathbf{I}\cdot \mathbf{S}(\phi)$, where $A$ is a constant,  $\mathbf{I}$ is the nuclear spin,  $\mathbf{S}(\phi) = \frac{1}{2}\sum_{ss'}c^{\dagger}_{\textbf{k}s}(\phi) \sigma_{ss'} c_{\textbf{k}s}(\phi)$ is the spin of an electron, and $c^{\dagger}_{\textbf{k}s}(\phi) = \cos(\phi) a^{\dagger}_{\textbf{k}s} + \sin(\phi) b^{\dagger}_{\textbf{k}s}$ is a linear combination of creation operators for the two {bands} 
with wavevector $\textbf{k}$ and spin $s$.  This hyperfine interaction gives rise to scattering with more or less emphasis on electrons from either {band} with the parameter $\phi$ allowing control over this relative weight, as shown by the form of $c^{\dagger}_{\textbf{k}s}(\phi)$.

\emph{INT Model}: This particular model is defined by the BdG Hamiltonian: $H_{BdG} = \sum_\textbf{k} \Psi^{\dagger}_\textbf{k} H(\textbf{k}) \Psi_\textbf{k}$, where 
\begin{equation}
    H(\textbf{k}) = \begin{bmatrix}
			h_0(\textbf{k}) & \hat{\Delta} \\ \hat{\Delta}^{\dagger} & -h_0(\textbf{k})
		\end{bmatrix},
\end{equation}
$h_0(\textbf{k}) = \mathbbm{1}_2 \otimes \left(v_\textbf{k} \cdot \textbf{k} \mathbbm{1}_2 + \gamma_\textbf{k} \cdot \textbf{k} \tau_x \right)$ exhibits a Dirac dispersion with parameters $v_\textbf{k}$ and $\gamma_\textbf{k}$ as introduced in \cite{Quan2022}, the pairing potential $\Delta = (\textbf{d} \cdot \sigma) i\sigma_y \otimes i\tau_y$ where $\tau_i$ are the Pauli matrices associated with the $a/b$ {bands},
$\mathbf{d}$ describes the triplet gap parameter, and  $\Psi_\textbf{k} = \nonumber\left(a_{\textbf{k}\uparrow}, b_{\textbf{k}\uparrow},a_{\textbf{k}\downarrow},b_{\textbf{k}\downarrow},a^{\dagger}_{-\textbf{k}\uparrow}, b^{\dagger}_{-\textbf{k}\uparrow}, a^{\dagger}_{-\textbf{k}\downarrow}, b^{\dagger}_{-\textbf{k}\downarrow}\right)$. In this model, the dependence on $\textbf{k}$ simplifies to dependence only on the component of $\textbf{k}$ perpendicular to the Fermi surface, as seen in the dispersion relation and subsequent coherence factor expressions. It is also important to note that with reference to the analogous model introduced in \cite{Ghosh2020}, the parameter $s$ that introduces an inherent energy difference in the $a$ and $b$ {bands} 
has been set to $0$, and $\delta$ has been endowed with linear $\textbf{k}$ dependence to yield the aforementioned Dirac dispersion. Furthermore, the order parameter is non-unitary and the condensate has a finite spin polarization  $\mathbf{q} = 2i\mathbf{d}\times\mathbf{d}^*$ \cite{Ramires2022}.  We consider a few different choices of pairing vector $\textbf{d} = \Delta_0\boldsymbol{\eta}$, where $\boldsymbol{\eta}$ is a complex unit vector. These are (I) $\textbf{d}_a = \frac{\Delta_0}{\sqrt{1+a^2}}(a,i,0)$ with $\textbf{q} \sim \frac{a}{1+a^2}\hat{z}$, (II) $\textbf{d}_z = (0,0,1)$ with $\textbf{q} = 0$, (III) $\mathbf{d}^{xy}_{\theta} = {\Delta_0}(1 + i\cos\theta_{\eta}, i\sin\theta_{\eta},0)/{\sqrt{2}}$ with $\mathbf{q}\sim \sin\theta_\eta \hat{z}$, and (IV) $\mathbf{d}^{yz}_{\theta} ={\Delta_0}(0,i\sin\theta_{\eta}, 1 + i\cos\theta_{\eta})/{\sqrt{2}}$ with $\mathbf{q}\sim \sin\theta_\eta \hat{x}$. $\theta_{\eta}$ characterizes the relative angle between the real and imaginary parts of $\mathbf{d}$ and the degree of non-unitarity. \VT{The first two cases correspond to the symmetry allowed irreducible representations for a single-band with weak spin orbit coupling \cite{Annett1990AP}.} Here, by increasing $a$ away from 0, the effect is to introduce TRSB with a nonzero $\boldsymbol{\omega} = i(\boldsymbol{\eta}\times \boldsymbol{\eta}*)$. 
There are two superconducting gaps, $\Delta_{1,2}$ given by: 
\begin{equation}
    \label{eqn:gap}
    \Delta_{1,2} = \Delta_0\sqrt{1-\left({\gamma_k}/{v_k}\right)^2}\sqrt{1\pm |\boldsymbol{\omega}|}.
\end{equation}

Evidently, having a complex order parameter with $a\neq0$ or real and imaginary parts that are not parallel is necessary for this model to exhibit two different gaps. As shown in the supplemental materials,  there are two types of Bugoliubov excitations:
\begin{eqnarray}
\alpha_{\textbf{k}s} &=& v_{\textbf{k}s}(a_{\textbf{k}s} - b_{\textbf{k}s}) - u^*_{\textbf{k}s} (a^{\dagger}_{-\textbf{k}s} + b^{\dagger}_{-\textbf{k}s}) \\ \nonumber \beta^{\dagger}_{-\textbf{k}s} &=& u_{\textbf{k}s}(a_{\textbf{k}s} - b_{\textbf{k}s}) + v_{\textbf{k}s}(a^{\dagger}_{-\textbf{k}s} + b^{\dagger}_{-\textbf{k}s})
\nonumber ,
\end{eqnarray}
where $u_{\textbf{k}s}$ and $v_{\textbf{k}s}$ are determined by the eigenvectors of $\mathcal{H}_{BdG}$ \cite{supp}. From these operators, one can determine a superconducting ground state that features pairing of like spins, in a given basis, and this is made possible by the inter-{band}
pairing allowed in the BdG Hamiltonian.

\emph{2-gap Singlet Model}: We also consider an effective model defined by inter-{band} pairing but which features singlet as opposed to triplet spin-pairing, i.e. a two-gap BCS model as in \cite{Sundar2024}. The BdG Hamiltonian matrix is similar to its INT counterpart, however the pairing potential now reflects the singlet nature of the electron pairing: $\Delta = -\sigma_y \otimes \Delta_0 \left(\cos(\theta_b)\mathbbm{1}_2 + \sin(\theta_b)\tau_x\right)$. Here the parameter $\theta_b$ is responsible for the existence of two distinct superconducting gaps $\Delta_{\pm}$ given by
\begin{align}
    \Delta_{\pm} = \Delta_0 \sqrt{1 \pm \sin(2\theta_b)}.
\end{align}
Similarly, this model can be solved to yield two species of spinful Bogoliubov excitations:
\begin{align}
    \alpha(\beta)_{\textbf{k}\uparrow} &= u_{\textbf{k}\alpha(\beta)} \left(a_{\textbf{k}\uparrow} \pm b_{\textbf{k}\uparrow}\right) + v_{\textbf{k}\alpha(\beta)}(a^{\dagger}_{-\textbf{k}\downarrow} \pm b^{\dagger}_{-\textbf{k}\downarrow}) \\ \nonumber \alpha(\beta)^{\dagger}_{-\textbf{k}\downarrow} &= -u^*_{\textbf{k}\alpha(\beta)} (a^{\dagger}_{-\textbf{k}\downarrow} \pm b^{\dagger}_{-\textbf{k}\downarrow}) + v_{\textbf{k}\alpha(\beta)} (a_{\textbf{k}\uparrow} \pm b_{\textbf{k}\uparrow}),
\end{align}
where the $u_{\textbf{k}\gamma}$ and $v_{\textbf{k}\gamma}$ factors are determined from the diagonalization of the Hamiltonian. Aside from the difference in spin-pairing, which is singlet in this case, the superconducting state will feature inter-{band} pairing. 

\emph{Coherence effects}: The nuclear spins will scatter the quasiparticles $\alpha_{\textbf{k}'\uparrow}\rightarrow\alpha_{\textbf{k}\downarrow}$ and $\beta_{\textbf{k}'\uparrow}\rightarrow\beta_{\textbf{k}\downarrow}$, and the spin lattice relaxation rate is calculated using Fermi's Golden Rule:
\begin{eqnarray}
     \nonumber T_1^{-1} &\sim& \sum_{\substack{\textbf{k},\textbf{k'},s,s' \\ i,j = \alpha,\beta}} \abs{A_{is'\rightarrow\ js}(\textbf{k},\textbf{k'})}^2 g_{ij}(\textbf{k},\textbf{k'},s,s').
    \label{eqn:T1}
\end{eqnarray}
Here $A_{\alpha/\beta s'\rightarrow\alpha/\beta s}(\textbf{k},\textbf{k'})$ are the scattering amplitudes for the two types of quasiparticles, and $g_{ij}(\textbf{k},\textbf{k'},s,s') = f_{j,\textbf{k},s} (1 - f_{i,\textbf{k'},s'}) \delta(E_{i,\textbf{k'},s'} - E_{j,\textbf{k},s} - \hbar \omega_N)$, where $f_{\epsilon \textbf{k}s}$ is the Fermi-Dirac function for quasiparticles of energy $E_{\epsilon \textbf{k}s}$, and $\omega_N$ is the nuclear resonance frequency.  The sums over $\textbf{k}$ and $\textbf{k'}$ can be converted to integrals over energy using the density of states, and we allow $\Delta_0$ in Eq.~\ref{eqn:gap} to have the temperature dependence given by the solution to the self-consistency equation for the singlet BCS theory gap \cite{Coleman}, as described in the supplemental material \cite{supp}.

We first consider fits to the INT model, shown 
in Fig. \ref{Fig:T1inv}. In this case, the scattering rates between the quasiparticles are dramatically suppressed if the two gaps have different values because energy cannot be conserved in the process. 
The nuclear resonance frequency is {set} to $\hbar\omega_N/k_B T_c = {10^{-4}}$ and $10^{-6}$ {in the cases of $d_a$ and $d_{\theta}$, respectively}, with no additional broadening included. 
When the two gaps are equal, the model describes the peak very well. However, this case is unitary and has no TRSB. 
When TRSB is introduced and the two gaps are different, the results depend on whether the nuclear spin quantization axis is the same as the equal-spin pairing axis. If the two are the same, that is, $\mathbf{d}$ lies in the $xy$ plane, the spin-flip process has an immediate downturn in \slrr\ and there is no possibility of an HS peak. When the axis for equal spin pairing is away from the nuclear quantization axis, that is, $\mathbf{d}$ lies in the $yz$ plane, a weak HS peak arises due to certain scattering processes that do not require crossing from one gap to the other; see \cite{supp} for further discussion. 
By making the nuclear frequency anomalously small $\hbar\omega_N/k_B T_c = 10^{-12}$ (0.04 Hz), the peak can be enhanced to fit the data well, as shown in the supplemental materials \cite{supp}. However, this artificial enhancement of the peak is not credible as one expects other forms of damping to reduce, rather than enhance the peak.

Fig. \ref{Fig:T1inv} also shows fits to the singlet model, assuming that $\hbar\omega_N/k_B T_c = 10^{-4}$ and $ 10^{-2}$ respectively. {The quasiparticle scattering must conserve energy, which includes the Zeeman (or quadrupolar) energy of the nuclear spin flip.  The peak will diverge if this energy is neglected and there is no impurity broadening, but will be gradually suppressed as this energy difference between the initial and final quasiparticle energies is increased.} We estimate the ratio for the material to be $10^{-4}$. There is a clear coherence peak below $T_c$, but it is larger than the experimental data for $\hbar\omega_N/k_B T_c = 10^{-4}$ {(dashed cyan line)}.  
The coherence peak is often modeled by introducing a broadening term that tends to suppress the divergence \cite{maclaughlinSCreviewbook}. A similar suppression can arise by assuming a larger nuclear frequency, 
and the solid blue line is calculated assuming $\hbar\omega_N/k_B T_c = 10^{-2}$, two orders of magnitude larger. This assumption suppresses the peak and the BCS case clearly fits the data well. The fitted gap values at $T=0$, given in Table \ref{tab:gaps},  are comparable to those reported by previous heat capacity \cite{Weng2016} and magnetic susceptibility measurements \cite{Ghimire2024}, but somewhat closer to one another than those reported by recent $\mu$SR experiments \cite{Sundar2024}.  As shown in Fig. \ref{Fig:T1invlog}, these values of the gap capture the slope in the \slrr\ vs $1/T$ plots relatively well, except at the lowest temperatures. The saturation of the data at the lowest temperatures suggests small residual impurity scattering.

\begin{figure}
\centering
\includegraphics[width=\linewidth]{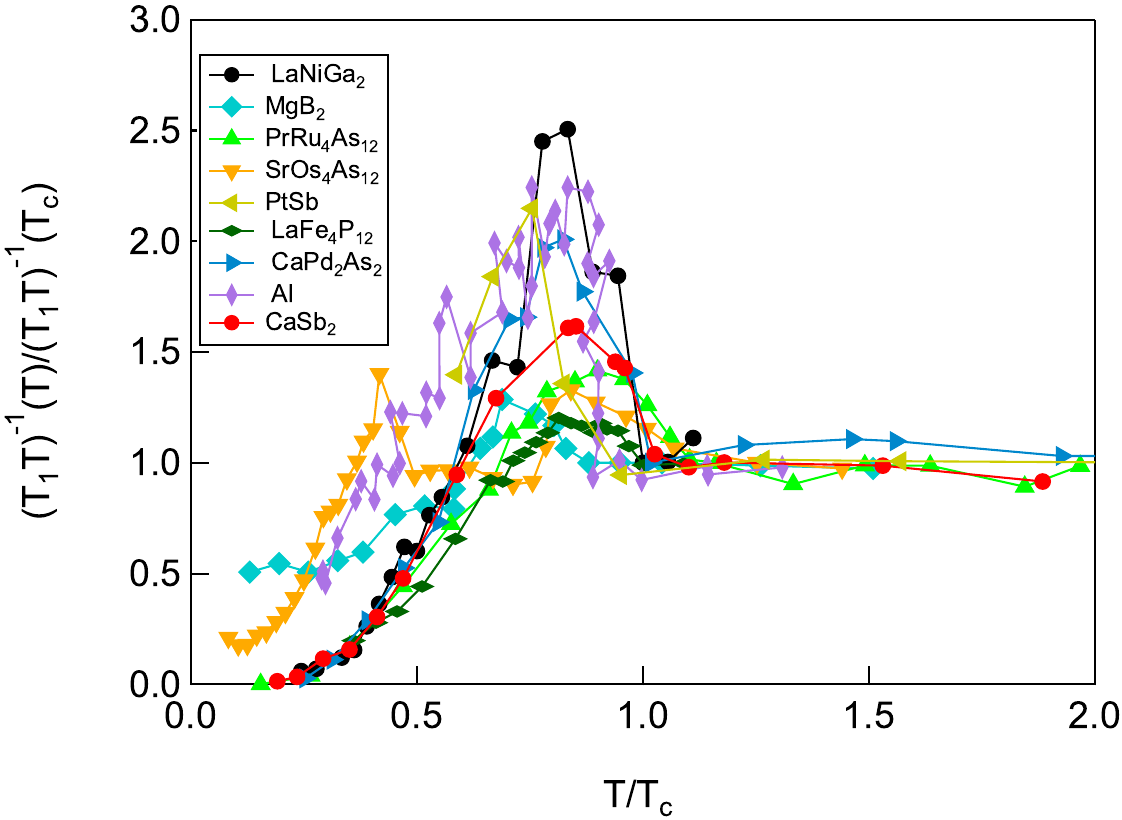}
\caption{\label{Fig:HSpeakCompare} 
Normalized $(T_1T)^{-1}$ versus reduced temperature for several different BCS superconductors, including: 
LaNiGa$_2$,  PrRu$_4$As$_{12}$ \cite{Shimizu2007}, 
CaPd$_2$As$_2$ \cite{Ding2016},
SrOs$_4$As$_{12}$ \cite{Ding2019},
PtSb \cite{Yamada2025},
LaFe$_4$P$_{12}$ \cite{IshidaLaFe4P12NMR},
MgB$_2$ \cite{MgB2NMR},
CaSb$_2$ \cite{Takahashi2021},
and Al \cite{masudaredfield}.}
\end{figure}

\begin{table}[b]
\caption{\label{tab:swave}Properties of selected s-wave superconductors. Note that measurements in LaFe$_4$P$_{12}$ and MgB$_2$ were carried out in an applied field (NMR), and the Al measurements were carried out in zero field via an unusual field-cycling procedure. All others were carried out in zero field (NQR).}
\begin{ruledtabular}
 \begin{tabular}{lllll}
        compound & $T_c$ (K) & $\Delta/k_BT_c$ & $\hbar \omega_N/k_B T_c$ $(\times10^{-3})$ & Ref \\ 
        \hline              
        PrRu$_4$As$_{12}$ & 2.3 & 1.6 & 1.18 & \cite{Shimizu2007} \\        
        LaRu$_4$As$_{12}$ & 10.3 & 1.85 & 2.64 & \cite{Shimizu2007} \\
        CaPd$_2$As$_2$ & 1.27 & 1.58 & 1.57 & \cite{Ding2016} \\ 
        SrOs$_4$As$_{12}$ & 4.8 & 1.25 & 0.62 & \cite{Ding2019} \\ 
        PtSb & 2.1 & 1.76 & 5.83 & \cite{Yamada2025} \\ 
        LaFe$_4$P$_{12}$ & 4.1 & 1.9 & 0.13 & \cite{IshidaLaFe4P12NMR} \\ 
        MgB$_2$ & 39.2 & 2.5 & 0.02 & \cite{MgB2NMR} \\ 
        Al & 1.178 & 1.6 & - & \cite{masudaredfield} \\ 
        CaSb$_2$ &  1.7 & 1.52 & 2.1  & \cite{Takahashi2021} \\
        LaNiGa$_2$ & 2.1 & 1.41, 1.95 & 0.11 & - \\
    \end{tabular}
    \end{ruledtabular}
\end{table}

The size of the coherence peak in LaNiGa$_2$, which reaches a factor of 2.5 times the normal state value, is large compared to many other superconductors, as illustrated in Fig. \ref{Fig:HSpeakCompare}.  The size of this peak may be partly related to the frequency at which the measurements were carried out.  
Table \ref{tab:swave} summarizes the superconductors shown in Fig. \ref{Fig:HSpeakCompare} and the normalized frequencies at which the experiments were carried out.  The NQR frequency of LaNiGa$_2$ is among the lowest measured. Overall, the robustness of the Hebel-Slichter peak in LaNiGa$_2$ compared to other materials suggests a conventional pairing.

In summary, the zero field NQR \slrrtext\ data for LaNiGa$_2$ are inconsistent with the INT model unless the two gaps are identical, in which case the state is unitary and there is no TRSB.  On the other hand, a two band BCS state with singlet pairing describes the data well.  This raises some doubts on the identification of triplet-pairing and time-reversal symmetry breaking in these materials.
Neither scenario is consistent with $\mu$SR measurements of TRSB.  However, $\mu$SR measurements are close to the lower limit of experimental detection and were conducted only on polycrystalline material. Further measurements should be performed on single crystals, and other measurements, such as optical Kerr rotation \cite{Schemm2014}
should be done to confirm the presence of TRSB. 
Measurements of the ultrasonic attenuation rate would provide further constraints on the pairing symmetry in LaNiGa$_2$ as that process involves a spin-independent relaxation and is complimentary to our NQR.

We acknowledge stimulating discussions with W. Pickett. Work at UC Davis was supported by the NSF under Grant No.  DMR-2210613, as well as the UC Laboratory Fees Research Program (LFR-20-653926). We acknowledge support from the Physics Liquid Helium Laboratory Fund. NQR measurements below $T_c$ were performed at Los Alamos and were supported by the U.S. Department of Energy, Office of Basic Energy Sciences, Division of Materials Science and Engineering project ``Quantum Fluctuations in Narrow-band Systems".

\bibliography{LaNiGa2bibliography}

\end{document}